# Unveiling pressurized bulk superconductivity in a trilayer nickelate $Pr_4Ni_3O_{10}$ single crystal

*Cuiying Pei[1]‡, Mingxin Zhang[1,2]‡, Di Peng[3,4]‡, Yang Shen[1]‡, Shangxiong Huangfu[5]‡, Shihao Zhu[1], Qi Wang[1,6], Juefei Wu[1], Junjie Wang[1], Zhenfang Xing[3,4], Lili Zhang[7], Hirokazu Kadobayashi[8], Saori I. Kawaguchi[8], Yulin Chen[1,6,9], Jinkui Zhao[2], Wenge Yang[3,4], Hongli Suo[5]*, Hanjie Guo[2]*, Qiaoshi Zeng[3,4]*, Guang-Ming Zhang[1,10]* and Yanpeng Qi[1,6,11]**

[1]School of Physical Science and Technology, ShanghaiTech University, Shanghai 201210, China

[2]Songshan Lake Materials Laboratory, Dongguan 523808, Guangdong, China

[3]Center for High Pressure Science and Technology Advanced Research, Shanghai 201203, China

[4]Shanghai Key Laboratory of Material Frontiers Research in Extreme Environments (MFree), Institute for Shanghai Advanced Research in Physical Sciences (SHARPS), Shanghai 201203, China

[5]Key Laboratory of Advanced Functional Materials, Ministry of Education, College of Materials Science and Engineering, Beijing University of Technology, Beijing, 100124, China

[6]ShanghaiTech Laboratory for Topological Physics, ShanghaiTech University, Shanghai 201210, China

[7]Shanghai Synchrotron Radiation Facility, Shanghai Advanced Research Institute, Chinese Academy of Sciences, Shanghai 201203, China

[8]Japan Synchrotron Radiation Research Institute, Sayo-gun Hyogo 679-5198, Japan

[9]Department of Physics, Clarendon Laboratory, University of Oxford, Parks Road, Oxford OX1 3PU, UK

[10]Department of Physics, Tsinghua University, Beijing 100084, China

[11]Shanghai Key Laboratory of High-resolution Electron Microscopy, ShanghaiTech University, Shanghai 201210, China

The recent discovery of superconductivity in pressurized Ruddlesden-Popper (RP) nickelates has provided new perspectives on the mechanism of high-temperature superconductivity. Up to now, most experiments concentrated on the lanthanum-related RP phase, so the discovery of new superconducting RP nickelates is highly desirable to reveal their generality. Here we report that high-quality $Pr_4Ni_3O_{10}$ single crystal is grown with an optical floating zone furnace under high oxygen pressure. High-pressure transport measurements show that the superconducting state arises above 10 GPa, and the maximum $T_c$ reaches 39 K without saturation, significantly exceeding the value of 25~30 K of $La_4Ni_3O_{10}$. Ultrasensitive d.c. magnetic susceptibility measurements under high pressure indicate bulk superconductivity with appreciable superconducting volume fractions. By performing *in situ* high-pressure synchrotron X-ray diffraction measurements at 16 K, a structural transition is found from monoclinic $P2_1/a$ to tetragonal $I4/mmm$. Unlike $La_4Ni_3O_{10}$, the electronic structure of the high-pressure phase of $Pr_4Ni_3O_{10}$ from density functional theory exhibits a dramatic metallization of the σ-bonding band consisting of three $d_{z^2}$ orbitals and van Hove singularity of coupled bands of $d_{x^2-y^2}$ orbitals near the Fermi level, similar to the bilayer nickelate $La_3Ni_2O_7$. These findings reveal some generic features of both crystal and electronic structures for high-temperature superconductivity in nickelates and multi-layer cuprates.

Superconductors with high superconducting transition temperatures ($T_c$) remain a long-sought goal within the field of condensed-matter physics. In addition to cuprates[1-5] and hydrides[6-11], $La_3Ni_2O_7$ is one of the few superconductors with $T_c$ above the liquid nitrogen temperature[12], which has opened up the era of nickel-based high-temperature superconductivity[13-32]. The recent observation of superconductivity at 82.5 K was reported in a Pr-doped $La_2PrNi_2O_7$ polycrystal under high pressure, accompanied by distinct diamagnetic signals[14]. The introduction of Pr has not only promoted the value of $T_c$ but also significantly enhanced the superconducting phase purity.

However, increasing the doping concentration of rare-earth elements in $La_{3-x}Pr_xNi_2O_7$ remains experimentally challenging. Fortunately, the trilayer Ruddlesden-Popper (RP) nickelates $Pr_4Ni_3O_{10}$ can be synthesized with high quality due to its efficient suppress of the intergrowth of different Ruddlesden-Popper phase during Pr substitution for La[33]. In 2020, the exploration of Pr-based nickel oxides was initiated with the successful growth of $Pr_4Ni_3O_{10}$ single crystals under oxygen pressure of 140 bar[34,35]. A distinctly anisotropic behavior between in-plane and out-of-plane was ascribed as well[34]. Synchrotron x-ray diffraction and neutron diffraction results identify and characterize the intertwined charge- and spin-density waves (CDW and SDW) on the Ni sublattice in $R_4Ni_3O_{10}$ (R = La, Pr)[35,36].

Inspired by the above work, a question arises naturally: is it possible to achieve high-temperature superconductivity in Pr-based trilayer RP nickelates? To address this, we synthesized single crystals using an optical-image floating zone furnace. The characterization of $Pr_4Ni_3O_{10}$ in ambient conditions confirmed its high quality. A series of transport measurements under high pressure was carried out with solid-, liquid-, and gas-state pressure-transmitting media (PTMs). It demonstrated that the subtle high-pressure structural phase of $Pr_4Ni_3O_{10}$ necessitates an exceptionally high degree of pressure hydrostaticity. The density wave (DW) in $Pr_4Ni_3O_{10}$ single crystal was suppressed by pressure, and subsequently zero resistance emerges at ~30 GPa with Helium as PTM. Within the scope of our research, a maximum and unsaturated $T_c$ of 39 K was achieved. A strong diamagnetic response confirmed the superconductivity of $Pr_4Ni_3O_{10}$, yielding a superconducting volume fraction exceeding 80%. Concurrently, a structural phase transition from monoclinic to tetragonal structure occurred. Assessing the evolution of pressure-induced $NiO_6$ octahedral configuration, which is enabled by synchrotron X-ray diffraction (XRD) measurements under high pressure and low temperature T=16 K, supports for the meticulous depiction of electronic structure under high pressure. The established correlation between crystal structure and electronic structure in trilayer nickelates is crucial for elucidating the superconducting mechanism of nickelates and designing novel high-temperature superconductivity materials.

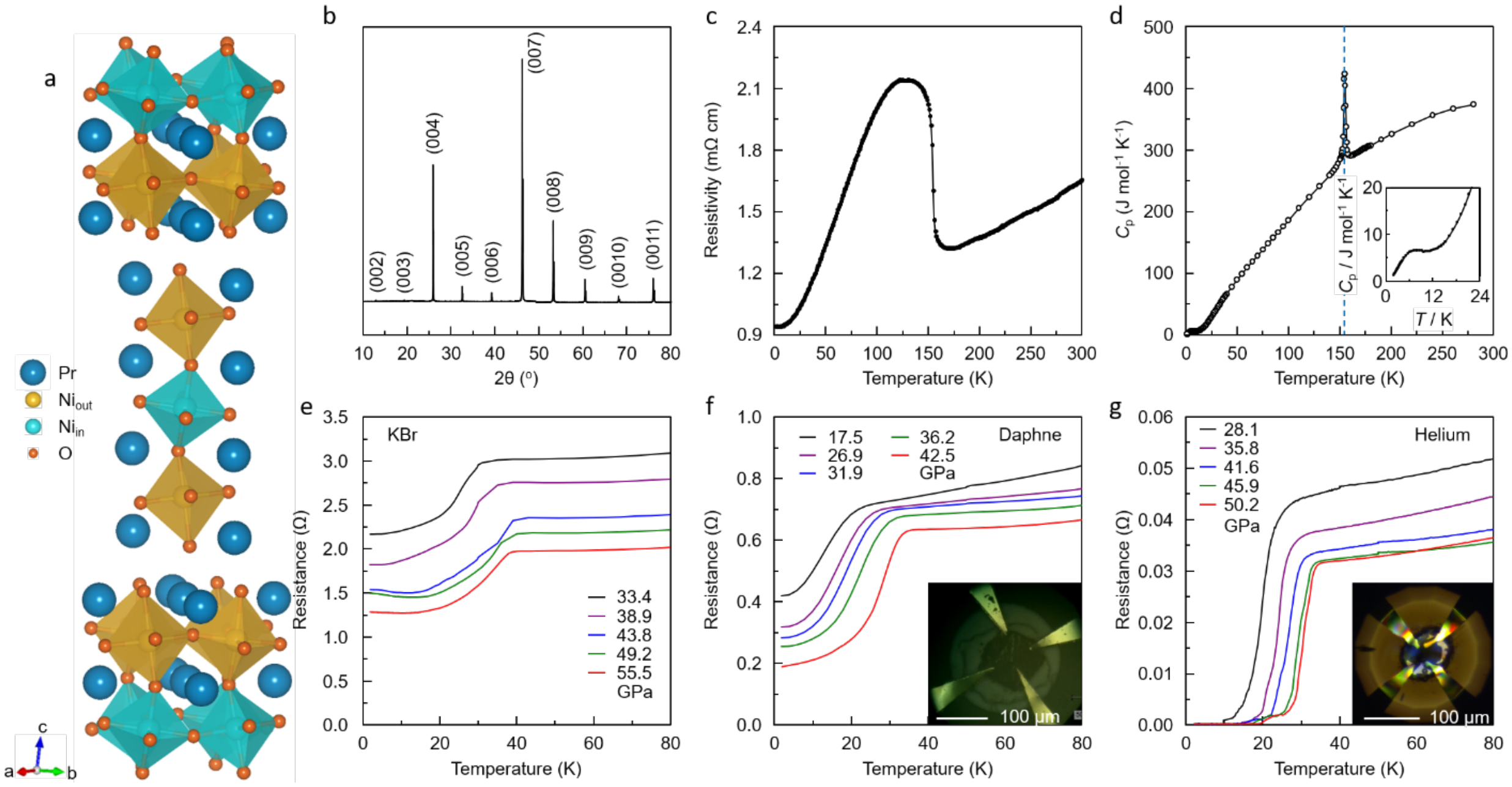

**Figure 1.** Characterization of $Pr_4Ni_3O_{10}$ at ambient pressure and temperature-dependent resistances under various pressures. a, Crystal structure of $Pr_4Ni_3O_{10}$. b, The room temperature XRD peaks from the ab plane of $Pr_4Ni_3O_{10}$ single crystal. c, Temperature-dependence of ab-plane resistivity $\rho(T)$ of $Pr_4Ni_3O_{10}$ single crystal. d, The heat capacity $C_p$ of $Pr_4Ni_3O_{10}$ single crystal, and the inset is the enlarged heat capacity $C_p$ in low temperatures. e, Enlarged $R(T)$ curves in the vicinity of the superconducting transition with KBr as PTM. f, Enlarged electrical resistance $R(T)$ of $Pr_4Ni_3O_{10}$ under different pressures with Daphne 7373 as PTM. The inset is a photograph of the electrodes used for high-pressure resistance measurements. g, Enlarged $R(T)$ curves in the vicinity of the superconducting transition with Helium as PTM. The inset is the photograph of the electrodes used for high-pressure resistance measurements with Helium as PTM.

$Pr_4Ni_3O_{10}$ is a member of the Ruddlesden-Popper trilayer rare-earth-nickelates, which is constructed by stacking of perovskite $(PrNiO_3)_3$ blocks and rock salt PrO layers along the *c* axis (Figure 1a). It crystallizes in the space group $P2_1/a$ with distorted $NiO_6$ octahedra. As reported, $Pr_4Ni_3O_{10}$ nominally contains $Ni^{2+}/Ni^{3+}$, with an average valence state of +2.67[35], which gives rise to strongly coupled electronic and magnetic phases. The $Pr_4Ni_3O_{10}$ single crystal was successfully grown using an optical floating zone furnace under an oxygen pressure of 140-150 bar[34,35]. Figure 1b shows the XRD peaks from the *ab* plane of $Pr_4Ni_3O_{10}$ single crystals at room temperature. All peaks can be well indexed to the (00l) reflections of $Pr_4Ni_3O_{10}$. The compositions derived from energy dispersive spectroscopy (EDS) measurement reveal an atomic ratio of Pr: Ni= 4: 3.09 (Figure S1). The thermogravimetric analysis (TGA) data revealed the oxygen content of the obtained crystal is about 9.99. The XRD, EDS and TGA results collectively confirm the high quality of $Pr_4Ni_3O_{10}$ single crystals.

The temperature-dependent resistivity of $Pr_4Ni_3O_{10}$ single crystal is depicted in Figure 1c. The resistivity decreases with decreasing temperature, displaying a metallic behavior. A distinct step-like kink is observed at around 156 K. This anomaly resembles the situation in $La_4Ni_3O_{10}$, suggesting the emergence of a DW state, as previously demonstrated in neutron diffraction experiments[37]. The heat capacity of $Pr_4Ni_3O_{10}$ is illustrated in Figure 1d, where a sharp peak at 156 K indicates a DW phase transition. The anomaly below 10 K could be attributed to a magnetic ordering from 4*f*-electrons of Pr ions[38].

The temperature-dependent resistance measurements of $Pr_4Ni_3O_{10}$ under high pressure were carried out with various PTMs. The DW suppression and subsequent metallization were clearly observed under increasing pressure using different PTMs, including no PTM, solid KBr, and liquid Daphne 7373. (Figure S2) Apparent drops in resistance (Figures 1e and 1f) were observed at ~20

GPa, however, the zero-resistance state was not achieved in any PTM cases. Due to the extreme sensitivity of superconducting properties of $Pr_4Ni_3O_{10}$ to pressure conditions, we employed Helium gas as the PTM to further improve the hydrostatic environment. As shown in Figure 1g, the resistance of $Pr_4Ni_3O_{10}$ exhibited metallic behavior with the pressure over 28.1 GPa and the resistance decreased as pressure increased. An obvious decrease in resistance emerged below 20 K at 28.1 GPa, and as the pressure increased it became more pronounced. At 35.8 GPa, the resistance $R(T)$ reached to zero around 15 K with an onset temperature of 26 K.

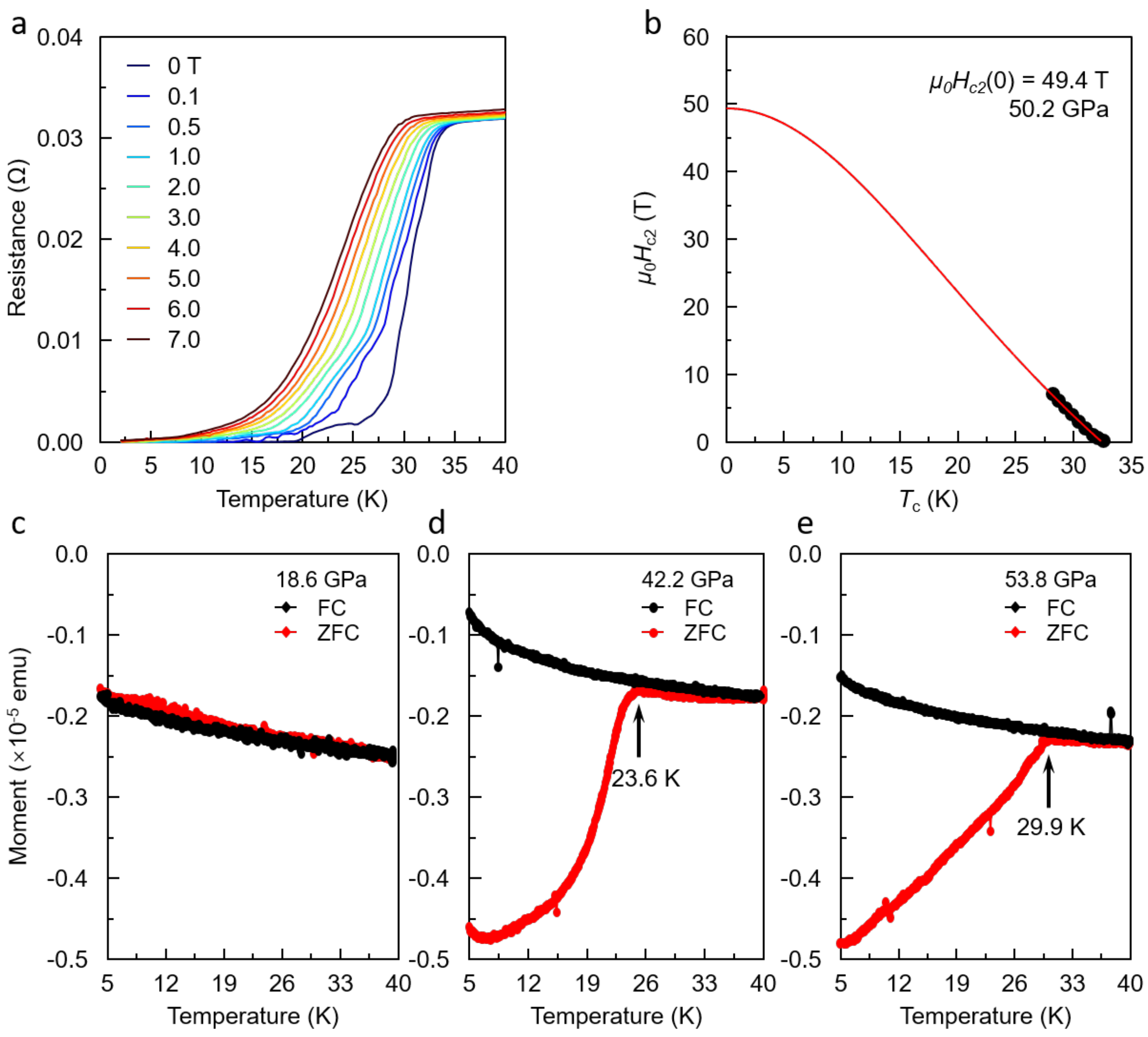

**Figure 2.** Resistance of $Pr_4Ni_3O_{10}$ in magnetic field and d. c. susceptibilities under various pressures. a, Field dependences of electrical resistance for sample $Pr_4Ni_3O_{10}$ at 50.2 GPa with Helium as PTM. b, Temperature dependence of the upper critical field $\mu_0H_{c2}(T)$ at 50.2 GPa with Helium as PTM. Here the values of $T_c$ are determined at 90% of the normal state resistance just above the onset superconducting transition temperature. The red solid line represents the fits using the Ginzburg-Landau formula. c-e, Temperature-dependent magnetization of $Pr_4Ni_3O_{10}$ under a magnetic field of 20 Oe using the zero-field-cooled (ZFC) and field-cooled (FC) modes under 18.6 GPa, 42.2 GPa and 53.8 GPa, respectively. Neon is used as PTM.

Moreover, we performed electrical resistance measurements under various magnetic fields at 50.2 GPa with Helium as PTM. As illustrated in Figure 2a, the resistance drop was progressively suppressed under external magnetic fields, and is shifted to approximately 29 K at 7 T. Then we estimated the upper critical field $\mu_0H_{c2}(T)$ by fitting it with the empirical Ginzburg-Landau formula $\mu_0H_{c2}(T) = \mu_0H_{c2}(0)(1 - t^2)/(1 + t^2)$, where $t = T/T_c$. (Figure 2b) The fitted zero-temperature upper critical field $\mu_0H_{c2}(0)$ of $Pr_4Ni_3O_{10}$, derived from the 90% resistance of the normal state around $T_c$, reached 49.4 T at 50.2 GPa. It should be noted that the $\mu_0H_{c2}(0)$ obtained exceeds the values reported for $La_4Ni_3O_{10}$, below the Pauli paramagnetic limit $H_P = 1.84T_c$. According to the relationship $\mu_0H_{c2} = \Phi_0/(2\pi\xi^2)$, where $\Phi_0 = 2.07 \times 10^{-15}$ Wb, the Ginzburg-Landau coherence length $\xi_{GL}(0)$ is given by 2.58 nm at 50.2 GPa.

To further determine the nature of the pressure-induced zero-resistance state, we conducted ultrasensitive direct current (d.c.) magnetic susceptibility measurements under high pressures using a custom-built miniature beryllium-copper alloy DAC with Neon as PTM. Systematic analysis of the high-pressure d.c. susceptibility data revealed no obvious transition in either the zero-field-cooled (ZFC) or field-cooled (FC) modes at 18.6 GPa. (Figures 2c, S3a and S3d) However, upon increasing the pressure to 42.2 GPa, a distinct diamagnetic transition was observed in the ZFC measurements (Figures 2d, S3b and S3e), which provides direct evidence of superconductivity under pressure. Thus, we estimated the maximum superconducting volume fraction, which exceeds 80% at 42.2 GPa, confirming the bulk superconductivity in $Pr_4Ni_3O_{10}$. Upon further compression, the $T_c$ evidenced by diamagnetic response rises from 23.6 K at 42.2 GPa to 29.9 K at 53.8 GPa (Figures 2e, S3c and S3f). At these pressures, the superconducting transition temperature derived from magnetic measurements is consistent with the results obtained from electrical transport measurements. This inter-method agreement provides robust corroboration of the pressure-induced superconducting state in $Pr_4Ni_3O_{10}$.

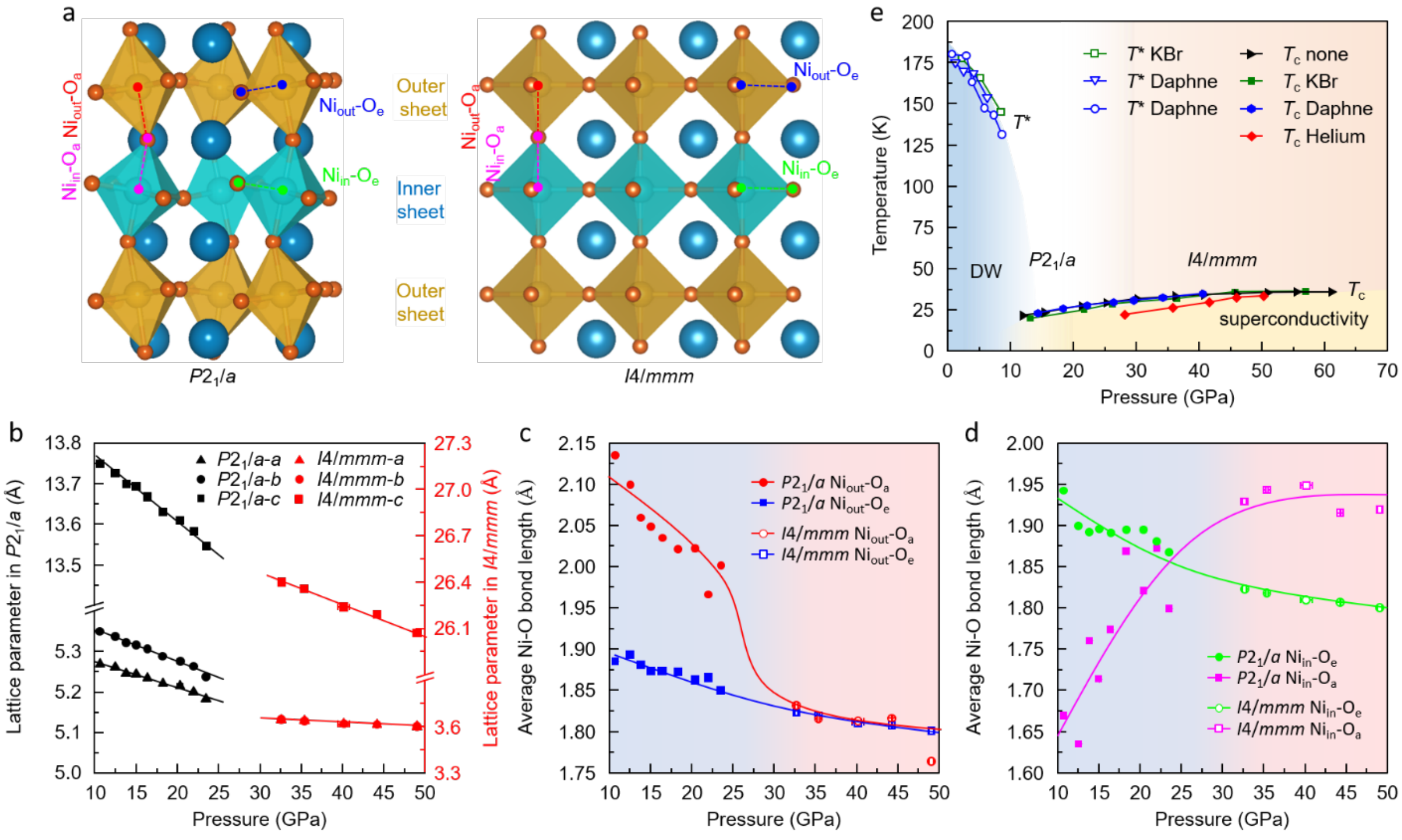

**Figure 3.** Pressure-dependent crystal structure and phase diagram of $Pr_4Ni_3O_{10}$. a, The schematic of two different kinds of disordered $NiO_6$ octahedra for $Pr_4Ni_3O_{10}$ in the low-pressure $P2_1/a$ phase and high-pressure $I4/mmm$ phase, respectively. b, Pressure dependence of lattice parameters $a$, $b$ and $c$ for $Pr_4Ni_3O_{10}$ in the $P2_1/a$ phase (black) and $a$, $c$ in $I4/mmm$ (red) phase extracted from the synchrotron XRD results obtained at 16 K. c, Average Ni-O bond length on equatorial and axial plane of $NiO_6$ octahedra in the outer perovskite layers as a function of pressure at 16 K. d, Average

Ni-O bond length on equatorial and axial plane of $NiO_6$ octahedra in the inner perovskite layers as a function of pressure at 16 K. e, Pressure dependence of the crystal structure, superconducting transition temperatures $T_c$ and DW transition temperature $T^*$ for $Pr_4Ni_3O_{10}$ as the function of pressure.

After establishing the pressurized bulk superconductivity in $Pr_4Ni_3O_{10}$ single crystals, it had better to determine the crystal and electronic structures under the conditions that the samples become superconducting. So, we first performed *in situ* synchrotron XRD measurements on powdered $Pr_4Ni_3O_{10}$ single crystals under various pressures at 16 K. (Figure S4) The Rietveld refinement based on monoclinic structure with $P2_1/a$ space group displays that the lattice parameters $a$ and $b$ are getting close to each other with increasing compression (Figures 3b and S5a), the ratio of $b/a$ drops dramatically and becomes metrically unity above ~30 GPa. The lattice parameters of $Pr_4Ni_3O_{10}$ under various pressures are smaller than those of $La_4Ni_3O_{10}$[39], suggesting that the substitution of Pr with La caused more chemical pre-compression effect (Figure S6). In addition, a clear volume discontinuity was observed with a volume drop of 3.0% when the pressure increases from 23.5 GPa to 32.7 GPa (Figure S7). All these observations suggest pressure-induced structural transition towards a higher symmetry structure, which is similar to that observed in $La_4Ni_3O_{10}$[37,40]. A typical Rietveld refinement based on tetragonal structure with space group $I4/mmm$ for the XRD pattern at 35.4 GPa and 16 K is shown in Figure S5b. Notably, the difference between $P2_1/a$ and $I4/mmm$ phase is too subtle to distinguish the pressure-induced structural phase transition when the synchrotron XRD was investigated at room temperature (Figure 8).

As depicted in Figure 3a, both of the low-pressure and high-pressure phases of $Pr_4Ni_3O_{10}$ contain two kinds layers consisting of $NiO_6$ octahedra: one inner layer and two outer layers. We examined the length of Ni-O bonds along the $c$-axis and within the $ab$ plane of the $NiO_6$ octahedra in both inner and outer layers. In the outer layers, the Ni-O bonds along the $c$-axis have a sudden decline after ~10 GPa in the $P2_1/a$ phase, which becomes a relatively constant ~1.83 Å in the $I4/mmm$ phase around 35 GPa (Figure 3c). The bond length in the *ab* plane also tends to ~1.83 Å in the $I4/mmm$ phase in the outer layers around 35 GPa, indicating the $NiO_6$ octahedra in the outer layers become more regular under high pressure. As for the inner layers, the Ni-O bonds along the *c*-axis have an abnormal incline in the $P2_1/a$ phase with increasing pressure, which converges to ~1.93 Å in the $I4/mmm$ phase around 35 GPa (Figure 3d). The Ni-O bonds in the *ab* plane of the inner layers share analogous features to those of outer layers, which become ~1.83 Å in the $I4/mmm$ phase around 35 GPa as well. This indicates the $NiO_6$ octahedra in the inner layers are less regular than those in outer layers. The population standard deviation $\sigma_{JT}$ in these two phases, which is used to scale the degree of deviation from the mean value, also confirms the improvement of the regularity of the Ni-O octahedra after the structure phase transition (Figure S9). Therefore, the atomic environment becomes more similar in both inner and outer layers after the structural transition, which induces relatively drastic variation along the *c*-axis. Such a process could improve the symmetry of crystals, conducive to the pairing of the superconductivity.

Notably, $NiO_6$ octahedra tilt and rotate during the pressure-induced phase transition. Additionally, the oxygen positions gradually shift from off-center positions between $Ni_{out}$-$Ni_{in}$ centers towards their central positions. The resulting periodic charge density distribution may account for the origin of the weak charge/magnetic order. Upon phase transition, the octahedral structure converges, and the oxygen positions approach the $Ni_{out}$-$Ni_{in}$ centers, leading to the DW disappearance. This is consistent with the DW suppression in electrical transport results under high pressure.

Based on the above electrical transport and structural results under pressure, the *T-P* phase diagram is summarized in Figure 3e, where the data are included from high-pressure measurements using a DAC, both in the absence of a PTM and with PTMs including a solid (KBr), a liquid (Daphne 7373), and a gaseous medium (Helium). In the lower pressure region, $Pr_4Ni_3O_{10}$ undergoes two transitions, forming the intertwined DW on the Ni sublattice and the magnetic order on the Pr sublattice. The DW state manifests as an anomaly in resistance, which is suppressed under pressure. A superconducting state emerges between 10-20 GPa. The value of $T_c$ increases dramatically with pressure, and a maximum $T_c$ of 39 K was obtained without saturation. The *in situ* high-pressure low-temperature XRD results indicate that the structural transition is a prerequisite to trigger superconductivity in $Pr_4Ni_3O_{10}$. Furthermore, the realignment of the $NiO_6$ octahedra and the chemical pre-compression by Pr atoms could be beneficial to the increase in $T_c$ under high pressure.

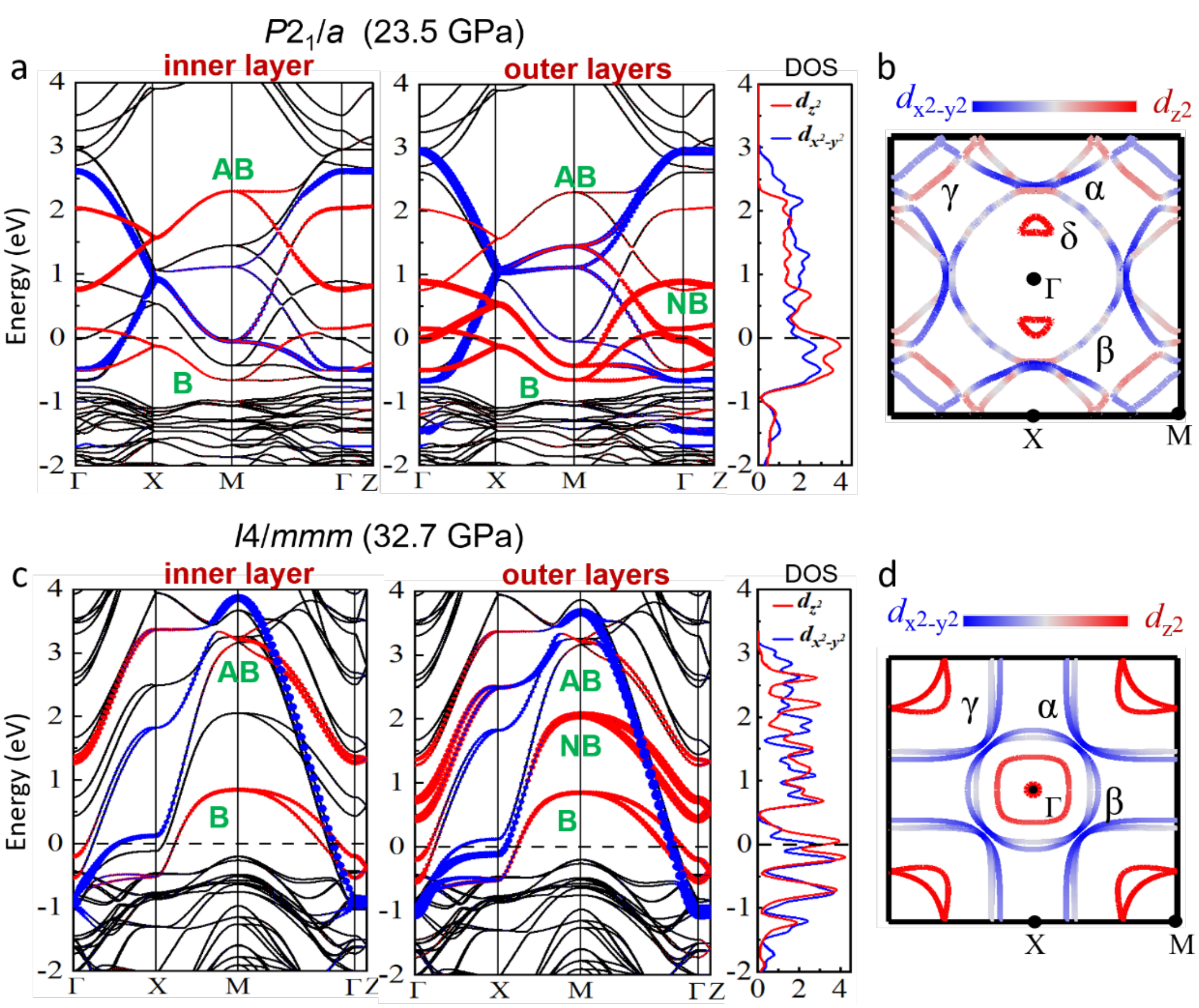

**Figure 4.** Layer and orbital resolved non-magnetic electronic structures of $Pr_4Ni_3O_{10}$ at hydrostatic pressure 23.5 GPa (low-symmetry $P2_1/a$) and 32.7 GPa (high-symmetry $I4/mmm$). a, c, Band structures with highlighted $d_{z2}$ (in red) and $d_{x^2-y^2}$ (in blue) orbital characters for the Ni inner and outer layers. We label the Ni $d_{z2}$ bonding (B), non-bonding (NB), and anti-bonding (AB) bands within the molecular orbital description. The tetragonal setting is used for the high symmetry $K$ point path. We adopt the notation of $d_{x2-y2}$ based on the local coordination of Ni-O octahedra, though monoclinic $P2_1/a$ structure is close to $\sqrt{2}\times\sqrt{2}$ supercell of the $I4/mmm$ structure with pronounced tilting and nearly 45° rotation of Ni-O octahedra. The corresponding total density of states for these orbitals are shown on the right. The Fermi level is marked by the horizontal dashed line. b, d, Fermi surfaces in the $k_z = 0$ plane with orbital-resolved α, β, and γ sheets.

To analyze the evolution of electronic structures of $Pr_4Ni_3O_{10}$ under external high pressure, we perform the calculation of density functional theory (DFT) methods for given typical pressures within the GGA+$U$ framework[41] based on Dudarev's method[42]. Here the on-site Coulomb interaction among the Ni electrons is chosen as $U$=4 eV, while the $f$-electronic states of Pr ions are pushed far away from the Fermi energy by using a sufficient large value of $U$ due to their nature of locality. As shown in Figure 4a, we can see that both the $d_{x2-y2}$ and $d_{z2}$ bands mixed with oxygen $p$-orbitals and distinguished their orbital contributions from inequivalent inner- and outer-layer Ni sites intersect with Fermi level, and are strongly inter-hybridized each other, being robust features under pressure. Due to the quantum confinement within the trilayer structural complex, the 3$d_{z2}$ orbitals strongly interact *via* inter-layer apical oxygen along the $c$ axis, forming three "molecular orbital bands". The molecular orbital description can be demonstrated in terms of a Ni trimer coupled with a nearest neighbor hopping parameter $t_\perp$, giving rise to the bonding (B), antibonding (AB) and non-bonding (NB) states[43,44]:

$$\varepsilon^{B}_{z2} = -\sqrt{\mathbf{2}}\boldsymbol{t}_{\perp},\ |\mathrm{B}\rangle = 1/2(1, \sqrt{\mathbf{2}}, 1),$$

$$\varepsilon^{NB}_{z2} = 0, \qquad |\mathrm{NB}\rangle = 1/\sqrt{\mathbf{2}}(1, \mathbf{0}, -1),$$

$$\varepsilon^{AB}{}_{z2} = \sqrt{2}t_{\perp}, \ |AB\rangle = 1/2(1, -\sqrt{2}, 1),$$

respectively. It can be seen that the B and AB states are formed by linear combinations of $d_{z2}$ orbitals from all three Ni layers with even parity, while NB state exclusively originates from a mixture of $d_{z2}$ orbitals from two outer Ni sites with odd parity. This interlayer molecular orbital scenario applies to all structures before and after the pressure-induced phase transition, as clearly demonstrated in layer- and orbital-resolved band structures in Figure 4a. Different from the $d_{z2}$ orbitals, the vertical interlayer coupling of the $d_{x2-y2}$ orbitals is much weaker, resulting in close energetics of $d_{x2-y2}$ bonding, non-bonding and anti-bonding bands.

Actually, a closer inspection can reveal that the Fermi surface of low-symmetry $P2_1/a$ phase comprises of several sheets (Figure 4b): hole-like α and δ pockets locate at the X point and near the Γ point, respectively, while the electron-like β and γ pockets center at the Γ and M points of the tetragonal Brillouin zone, respectively. Due to the monoclinic lattice distortion, the α, β and γ pockets have mixed characters of $e_g$ orbitals, while the δ pocket is predominantly in $d_{z2}$ character. With increasing pressure in the low-symmetry $P2_1/a$ phase, the band widths of both orbitals are increased and δ pocket is narrowed. (Figures 4a, 4b and S10)

However, after the structural phase transition, the Fermi surface topology is significantly changed, and electronic structure exhibits many new features (Figures 4c and 4d). 1) The Fermi surface consisting of one hole-like α sheet with dominant $d_{x2-y2}$ character is centered at M point, while an electron-like $\beta$ sheet with the mixed $e_g$ orbital character is around the Γ point. 2) Another hole-like γ pocket featured with dominant $d_{z2}$ bonding bands sits around the M point, and extra electron pockets from the $d_{z2}$ bonding band emerging at Γ point cross the Fermi level. 3) The $d_{x2-y2}$ orbital bands exhibit the van Hove singularity at the X point with peaked density of states near the Fermi level (Figure 4c). Crucially, there is a dramatic change of the γ pocket from electron-like to hole-like with the onset of the structural phase transition.

Our electronic structure calculations based on experimentally determined low-temperature crystal structures provide an opportunity to compare band filling and Fermi surface topology of $Pr_4Ni_3O_{10}$ with those of $La_4Ni_3O_{10}$ and $La_3Ni_2O_7$ superconducting phases. A formal valence count for $La_4Ni_3O_{10}$ and $Pr_4Ni_3O_{10}$ renders a $Ni^{2.67+}$ state corresponding to an average $3d^{7.33}$ filling of the Ni *3d* shell, while the valence count for $La_3Ni_2O_7$ gives a $Ni^{2.5+}$ state corresponding to an average $3d^{7.5}$ filling. Considering the fact that the large crystal field splits $t_{2g}$ orbitals away from the Fermi level, only four (three) electrons are active in the $e_g$ orbitals near the Fermi level for $La_4Ni_3O_{10}$ and $Pr_4Ni_3O_{10}$ ($La_3Ni_2O_7$), respectively. The way how to fill up these electrons in the $e_g$ orbitals imposed by strong interlayer coupling of $d_{z2}$ orbitals may crucially affect the nature of superconductivity.

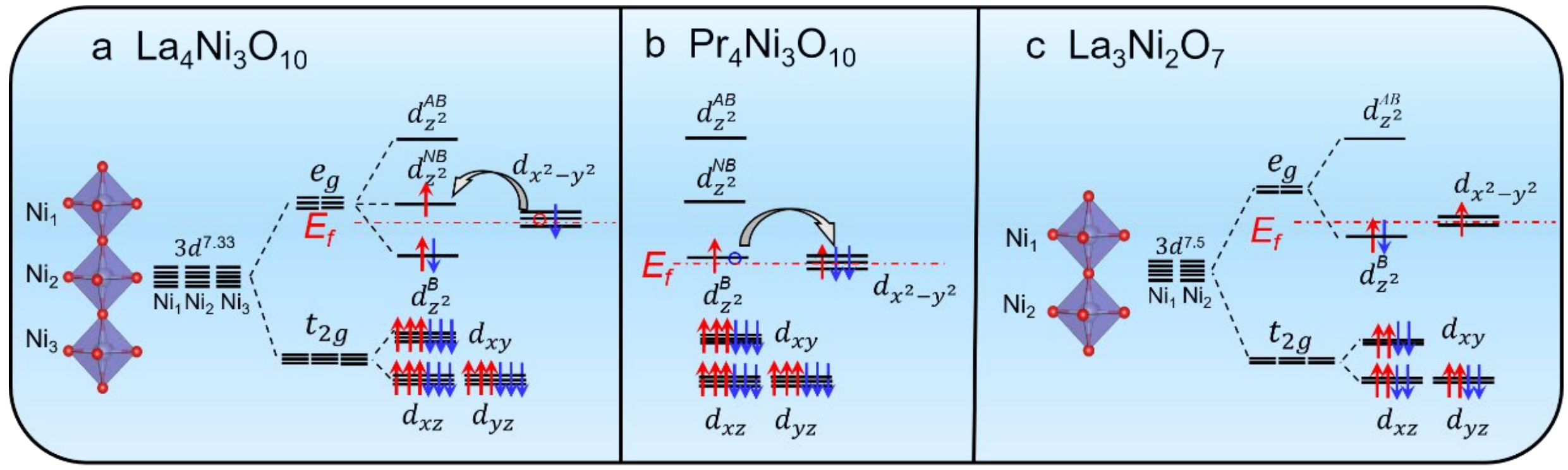

**Figure 5.** Simplified electronic energy levels of the nickel oxide complexes showing the similarities and differences among various RP superconducting nickelates a, $La_4Ni_3O_{10}$, b, $Pr_4Ni_3O_{10}$ and c, $La_3Ni_2O_7$. Due to quantum confinement of the layered structure, the molecular orbital representation of the Ni trimer is more natural than individual atomic orbitals of each layer for the description of $d_{z2}$ orbital electrons. Band fillings are terminated by the Fermi level $E_f$, which are extracted from band structures. Charge transfer between $d_{x2-y2}$ and $d_{z2}$ orbitals due to the energy

overlap is denoted by arrows. Notice that the energy splitting of $d_{x2-y2}$ orbital is less obvious as compared to the $d_{z2}$ orbitals, as the vertical interlayer coupling for $d_{x2-y2}$ orbitals is much smaller than that of $d_{z2}$ orbitals.

In Figure 5, we display the electronic energy levels of the nickel oxide complexes of $Pr_4Ni_3O_{10}$, $La_4Ni_3O_{10}$ and $La_3Ni_2O_7$ within the interlayer molecular orbital scenario. In the $La_4Ni_3O_{10}$ case, four electrons of $e_g$ orbitals occupy three different bands in the manner that the $d_{z2}$ bonding band just touches the Fermi surface (nearly fully filled), while the $d_{z2}$ non-bonding band and $d_{x2-y2}$ band cross the Fermi surface (partially filled), see Figure 5a. In contrast, four electrons in $Pr_4Ni_3O_{10}$ only occupies $d_{x2-y2}$ band (partially filled) and $d_{z2}$ bonding band (partially filled), as described in Figure 5b. Large proportion of electrons occupies $d_{x2-y2}$ band, leading to the enhanced metallization of the $d_{z2}$ bonding band with enlarged hole pockets centered at the M point. As a consequence, $Pr_4Ni_3O_{10}$ has an equivalent downward shift in the Fermi energy as compared to $La_4Ni_3O_{10}$. More importantly, $Pr_4Ni_3O_{10}$ has striking similarities with $La_3Ni_2O_7$, where the $d_{z2}$ bonding band slightly touches the Fermi level (nearly fully filled) and the $d_{x2-y2}$ band cross the Fermi surface (quarter-filled). Only $d_{x2-y2}$ and $d_{z2}$ bonding bands are active orbitals around the Fermi level, corresponding to a dramatic metallization of the σ-bonding band consisting of $d_{z2}$ orbitals within the trilayer unit, and an effective two-orbital description may be enough to describe the low-energy physics for both $Pr_4Ni_3O_{10}$ and $La_3Ni_2O_7$. The metallization of the inter-layer σ-bonding bands recalls the emergence of the conventional high-temperature superconductivity in $MgB_2$[45] and hydrides[6-8]. Therefore, the electronic structure characteristics of $Pr_4Ni_3O_{10}$ may lie between those of $La_4Ni_3O_{10}$ and $La_3Ni_2O_7$.

Furthermore, the trilayer $Pr_4Ni_3O_{10}$ superconductors also share the similar electronic structure and Fermi surface topology with multilayer high-$T_c$ cuprate superconductors[46,47], such as $YBa_2Cu_3O_{7-\delta}$ ($T_c$~93 K), $HgBa_2CaCu_2O_{6+\delta}$ ($T_c$~120 K), and $Tl_2Ba_2CaCu_2O_{8+\delta}$ ($T_c$~110 K). In these multi-layer cuprates, we can clearly find that two similar active electronic bands near the Fermi level originate from the bonding and anti-bonding $d_{x2-y2}$ orbital bands from the outer layers, while the inner layer if exists is not metallic. In particular, the van Hove singularity is also exhibited around X point. Such similitude between $Pr_4Ni_3O_{10}$ and multi-layer high-$T_c$ cuprates could make $Pr_4Ni_3O_{10}$ another candidate of superconductivity with unconventional pairing symmetry.

ASSOCIATED CONTENT

**Supporting Information**. Experimental section, energy dispersive spectroscopy spectrum in ambient condition, magnetic susceptibility in ambient condition, electrical resistivity under high pressure, magnetization under high pressure, XRD under high pressure and low temperature, typical Rietveld refinement, lattice parameter and volume evolution under high pressure, XRD under high pressure and room temperature, population standard deviation of Ni-O bond length, electronic band structure at 3.1 GPa are supplied as Supporting Information. This material is available free of charge *via* the Internet at http://pubs.acs.org.

AUTHOR INFORMATION

**Corresponding Author**

**Y.P.Q.** (qiyp@shanghaitech.edu.cn) **G.M.Z** (zhanggm@shanghaitech.edu.cn) **or Q.S.Z.** (zengqs@hpstar.ac.cn) **or H.J.G** (hjguo@sslab.org.cn) **or H.L.S** (honglisuo@bjut.edu.cn)

**Author Contributions**

Y.Q. designed the project. M.Z. and S.H. grew the single crystals with the support of H.S., H.G. and J.Z.; C.P., D.P., Z.X. and S.Z. performed the resistance measurements at varying pressures.

C.P., M.Z., Q.W. and J.J.W. performed the synchrotron XRD measurements with the support of L.Z., H.K., S.K. and Y.C.; D.P. and Z.X. conducted the high-pressure susceptibility measurements with the support of W.Y. and Q.Z. C.P. and M.Z. conducted the structural analysis. Y.S. and J.F.W. performed the DFT calculations. G.M.Z. proposed physical pictures to understand both the experimental and numerical results, and outlined the paper. C.P., G.M.Z. and Y.Q. wrote the paper with inputs from all co-authors.

‡These authors contributed equally.

**Notes**

While preparing this paper, we also noted three related studies that reported resistance measurements under high pressure on $Pr_4Ni_3O_{10}$ polycrystalline and single crystal samples[48-50].

ACKNOWLEDGMENT

This work was supported by the National Key R&D Program of China (Grant No. 2023YFA1607400) and the National Natural Science Foundation of China (Grant Nos. 52272265, 12474018). Q.S.Z., D.P. and W.Y. acknowledge the support from the National Natural Science Foundation of China (Grant Nos. 51871054, 52101187) and Shanghai Key Laboratory of Material Frontiers Research in Extreme Environments, China (Grant No. 22dz2260800), the Shanghai Science and Technology Committee, China (Grant No. 22JC1410300). H.J.G acknowledges the support from the National Natural Science Foundation of China (Grant No. 12004270) and Guangdong Basic and Applied Basic Research Foundation (Grant No. 2022B1515120020). S.H. acknowledges the Beijing Postdoctoral Research Foundation (2024-ZZ-023). H.L.S. acknowledges the National Natural Science Foundation of China (Grant No. 52277021). G.M.Z. acknowledges the support from the National Key Research and Development Program of MOST of China (Grant No. 2023YFA1406400). The authors thank the support from Analytical Instrumentation Center (# SPSTAIC10112914), SPST, ShanghaiTech University. The authors thank the staff members at BL15U1 in Shanghai Synchrotron Radiation Facility for assistance during data collection.